\documentclass[useASM, referee]{mn2e}
\input{psfig.sty}
\usepackage{graphicx}

\begin{document}
\title{Revisiting the characteristics of the spectral lags in short gamma-ray bursts \footnote{send offprint request to: z-b-zhang@163.com}}
\date{2005 Apr. 13}
\pubyear{????} \volume{????} \pagerange{2} \onecolumn

\author[Z. B. Zhang et al.]
       {Zhibin Zhang$^{1, 3}$, G. Z. Xie$^{1}$, J. G. Deng$^{2}$, W. Jin$^{1, 3}$\\
        $^1$Yunnan Observatory, National Astronomical Observatories, Chinese Academy of Sciences, \\
            P. O. Box 110, Kunming 650011, China\\
        $^2$Physics Department, Guangxi University, Nanning, Guangxi 530004, P. R. China\\
        $^3$The Graduate School of the Chinese Academy of Sciences\\
        }
\date{Accepted ????.
      Received ????;
      in original form 2006 April 10}

\pagerange{\pageref{firstpage}--\pageref{lastpage}} \pubyear{2001}

\maketitle
\label{firstpage}
\begin{abstract}
In this paper, we restudy the spectral lag features of short bright
gamma-ray bursts ($T_{90}<2.6$s) with a BATSE time-tagged event
(TTE) sample including 65 single pulse bursts. The cross-correlation
technique is adopted to measure the lags between two different
energy channels. Meanwhile, we also make an investigation of the
characteristics of the ratios between the spectral lag and the full
width at half maximum ($FWHM$) of the pulses, called relative
spectral lags (RSLs). We conclude that spectral lags of short gamma
ray bursts (SGRBs) are normally distributed and concentrated around
the value of 0.014, with 40 percent of them having negative lags.
With K-S tests, we find the lag distribution is identical with a
normal one caused by white noises, which indicates the lags of the
vast majority of SGRBs are so small that they are negligible or
nonmeasurable, as Norris \& Bonnell have suggested.

\end{abstract}

\section{Introduction}

The phenomena of SGRBs remained more mysterious than long ones until
recently, owing to a lack of more information about observations
such as afterglows. They have been found to be harder than long
bursts in spectra (e.g. Kouveliotou et al. 1993; Hurley et al.
1992). The spectra for both types generally evolve with time (Liang
\& Kargatis 1996; Band 1997). Most long GRB pulses reach their
maximum earlier in higher energy channels, called ``hard-to-soft''
evolution with positive spectral lags (Norris et al. 2000; Daigne \&
Mochkovitch 2003; Chen et al. 2005), while the issue of SGRBs on
spectral lags is still controversial, though some timescale analysis
techniques (Scargle 1998; Liang et al. 2002; Li et al. 2004) have
been employed to improve it.

The distribution characteristics of SGRB spectral lags had
previously been explored by several authors. We can divided their
viewpoints into two typical cases: one is that Norris et al. (2001,
2006) found the time lags in SGRBs were close to symmetric about
zero and confirmed they could be negligible in statistics. Another
is that Gupta et al. (2002) studied the lags with a sample of 156
SGRBs and found that one quarter of the total had negative lags,
implying ``soft-to-hard'' spectral evolution. Recently, using a
sample of 308 SGRBs, Yi et al. (2006) redid the analysis of time
lags and conclude that the sources with negative spectral lags are
in the minority (only about 17 percent of sources in their sample
locating at the side of negative lags). Prompted by the
contradictory results, we thus want to research what the features of
the spectral lags in SGRBs should in essence be. To avoid the
contamination of adjacent pulses by overlap, which could inevitably
produce additional errors owing to selection effects, we construct
our sample from SGRBs with single pulses in order to calculate the
unbiased lags accurately.

We have shown in our recent work (Zhang et al. 2006, hereafter paper
I) that the RSL of long bursts can be used as a good redshift
estimator. Since pulse widths and spectral lags are respectively
proportional to $\Gamma^{-2}$ approximately (Qin et al. 2004; Zhang
\& Qin 2005 and paper I) and the relationship of them with energy is
roughly $\sim E^{-0.4}$ (Fenimore et al. 1995; Norris et al. 1996),
such a definition of RSL can eliminate the influence of Lorentz
factor and energy on our analysis and thus the RSL could be regarded
as an intrinsic parameter. Considering this virtue, we expect to
know what the distribution of RSL should be. In $\S2$ we give the
data preparation; Definitions and measures of the physical variables
are shown in $\S3$; We list our new results in $\S4$. We end with a
brief conclusion and discussion in $\S5$.

\section{Data preparation}

\subsection{Sample selection}

Considering the duration attribute of SGRBs, the high-resolution TTE
data at 5 ms resolution are adopted to constitute our sample,
including 65 bursts. This sample consists only of single peaked
bursts. The motivation for this selection is that spectral lag
analysis is usually affected by both energy channel pair and time
resolution, as well as the neighbor pulses due to overlapping
effects. More detailed explanations for this can be found in paper
I. Consequently, the selection requirements in terms of the above
considerations are given as follows: $T_{90}$ duration $<$ 2.6 s;
BATSE peak flux (50-300 kev) $>$ 2 photons $cm^{-2} s^{-1}$; and
also peak count rate ($>$ 25 kev) $>$ 14000 counts $s^{-1}$.
Certainly, a bright-independent analysis is required as the
requirement for burst duration measurement (Bonnell et al. 1997) and
the signal-to-noise (S/N) levels are also taken into account, so
that we can take more relatively accurate measurements of such
variables as width and spectral lag of pulses.

\subsection{Background subtraction}

The first step is to select an appropriate background for
substraction. In terms of the existing opinion (e.g. McBreen et al.
2001), the median filter method is suitable for background
subtraction of SGRB data because of their special duration and
background level. Firstly, the start and end times in each pulse are
determined. The section involving pre- and post-pulse is generally
regarded as background. With the median method, the corresponding
median within the range of background data points is easily
estimated. Secondly, the estimated value of the background is
subtracted from the whole original data. According to this
methodology, we apply the same measurement to perform this analysis
of subtraction for all four energy channels (channel 1, 20-50 Kev;
channel 2, 50-100 Kev; channel 3, 100-300 Kev; channel 4, $>$300
Kev) respectively.

\subsection{Denoising}

After the background in each channel is subtracted the remaining
data are still fluctuating because of the disturbance of noise. In
theory, we must smooth these rough data in order to obtain the pure
signal data for the sake of analysis. For one burst, we hence adopt
the following model describing the pulses' shape (Kocevski et al.
2003) to fit the disturbed data
\begin{equation}
F(t)={F_m}(\frac{t}{t_m})^r[\frac{d}{d+r}+\frac{r}{d+r}(\frac{t}{t_m})^{(r+1)}]^{-\frac{r+d}{r+1}}
\end{equation}
where $t_m$ is the time of the maximum flux ($F_m$), of the pulse
and the quantities of $r$ and $d$ are two parameters depicting the
pulse shape. It has been shown that this function is very flexible
and powerful for describing a single peaked pulse.

We find from this model that there are four fitted parameters. In
fact, the exact function used to do the model fitting should include
an additional free parameter, $t_0$, denoting the start time of a
pulse. The application of this zero parameter is advantageous to our
analysis, because it can quicken the fit process and optimize the
results. Once we have finished the disposal of noise, the original
data are then refined into the usable signal data, which can be
utilized to calculate the spectral lags and widths of pulses, as
well as other physical quantities.

\section{Quantity definitions and measures}

\subsection{Spectral lag}

To get the spectral lag, we cross-correlate variation signals in
different energy channels $j$ and $k$ with the following
cross-correlation function (CCF) \cite{Ba97}
\begin{equation}
CCF(\tau; \upsilon_{j},
\upsilon_{k})=\frac{<\upsilon_{j}(t)\upsilon_{k}(t+\tau)>}{\sigma_{\upsilon_{j}}\sigma_{\upsilon_{k}}}
(j > k)
\end{equation}
where $\sigma_{\upsilon_{i}}=<\upsilon_{i}^{2}>^{1/2}$, and $i$
stands for the $i^{th}$ energy channel; $\tau$ is the general
spectral lag between energy bands $j$ and $k$, namely $\tau_{jk}$;
$\upsilon_{j}$ and $\upsilon_{k}$ represent two time series in which
they are respective light curves in two different energy bands.

The value of $\tau_{jk}$ is determined by the location of $\tau$
where CCF peaks, because the CCF curve caused by smoothed signal
data is considerably smooth and resembles a gaussian shape near its
peak on this occasion. If the data points around the peak are sparse
enough, we interpolate them in its neighborhood in order to reduce
the calculation errors.

\subsection{Relative spectral lag}

For long bursts, wider-pulse bursts had been found to preserve
longer spectral lags (Norris et al. 2005). This conclusion seems to
be universal not only for long bursts but also for short ones. In
addition, the same law about the relation of time lags to pulse
width also seems to coexist between distinct energy channels within
a burst. We thus are inspired to probe how their ratios evolve with
energy and other relevant parameters with the variety of spectral
lags and widths. For this purpose, unlike in paper I, we redefine
the quantity RSL as
\begin{equation}
\tau_{rel,jk}=\tau_{jk}/FWHM_{(k)} (j>k)
\end{equation}
where $FWHM_{(k)}$ denotes the full width at half maximum of the
time profile in the $k^{th}$ discriminator. From the definition, one
can see that $\tau_{rel, jk}$ is connected with the pair of energy
channels and is indeed a dimensionless quantity.

Here, the values of the subscript $k$ are assigned to be 1, while
another subscript $j$, denoting a higher energy channel, is taken as
$j=3$. Although $FWHM_{(k)}$ are usually dependent on $\tau_{jk}$,
the relationship doesn't influence our analysis on its credibility
(see paper I). If they are precisely measured, the corresponding
RSLs can be well determined within a certain significance level.

\section{Results}

After the subscripts in eqs. (2) and (3) are assigned as $k=1$ and
$j=3$, combining eqs. (1)-(3) we can then derive the quantities,
such as the spectral lag ($\tau_{31}$), the width ($FWHM_{(1)}$) and
the RSL ($\tau_{rel, 31}$). In the following section, we study the
properties associated with the above quantities of SGRBs, and list
our results comparatively.

\subsection{The distribution of $\tau_{31}$}

In the case of single peaked SGRBs, we make a plot in figure 1
\begin{figure} \centering
 \includegraphics[width=4.2in,angle=0]{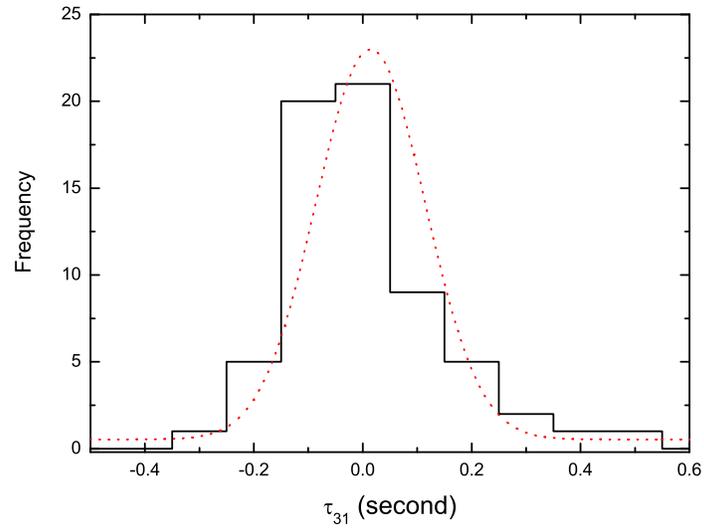}
    \caption{Distribution of spectral lags for a sample including 65 single peaked SGRBs.
    The dotted line represents the best fit to the distribution with a gaussian function, where the expectation value
     and the standard deviation are $\mu=0.014$ sec and $\sigma=0.1$ sec respectively. Fewer sources (40\%) with negative
     lags, in contrast with those with positive lags, have a short tail of the population lying to the left-hand side of zero.}
  \label{fig1}
 \end{figure}in order to gain the spectral lag distribution, from which we find
the lags have a distribution which is close to symmetric about zero.
The distribution looks like a gaussian function outline. In order to
prove this, we try to fit the distribution with a gaussian model and
obtain $\chi^{2}/dof=1.5$ with $R^{2}=0.98$, which indicates that
the spectral lags are indeed normally distributed with a high
confidence. Additionally, we find from figure 1 that the number
($\sim$ 40\%) of SGRBs with negative lags are slightly less than
those with positive values.

It is well known that white noise (random errors) can also lead to a
normal distribution with the expectation value $\mu=0$. Thus, our
task in this situation is to check if the gaussian distribution of
the spectral lags is in agreement with the distribution caused by
noise taken as $\mu=0$ and $\sigma=0.1$.
\begin{figure}
\centering
   \includegraphics[width=4.2in,angle=0]{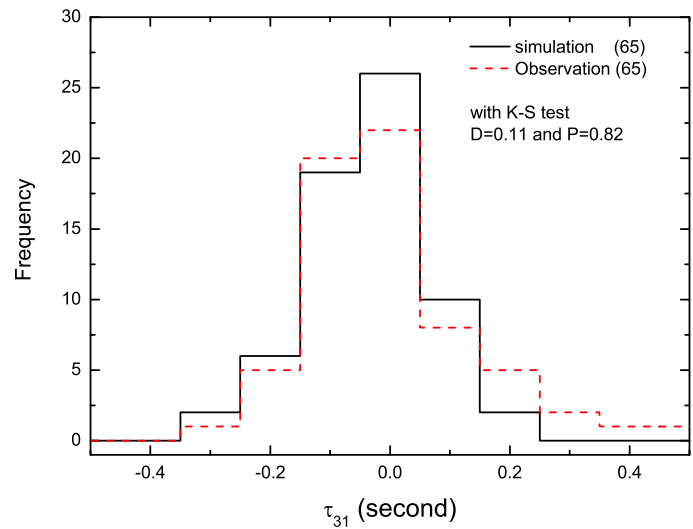}
  \caption{Comparison between the distribution of spectral lags in SGRBs and that of white
  noises. The sample size for both of them is equally 65. The former is
  marked with dashed line and the latter is identified by solid line. Symbols have been denoted in this plot.}
  \label{Fig2}
\end{figure}
Figure 2 shows the comparison between them. We apply a general K-S
test to the two one-dimensional distributions. Surprisingly we
discover that they are in excellent agreement with each other, with
large confidence probability $P\sim0.82$ and small K-S statistic
$D=0.11$. Furthermore, we find there are about 94\% of the total
sources which have zero lag within a $3\sigma$ fiducial limit. These
results highly support the conclusion of a larger majority (90-95\%)
with nought lag drawn by Norris \& Bonnell (2006).

The coincidental results demonstrate that the spectral lags of SGRBs
are so small that they are comparable with random errors. The reason
the lags follow such a normal distribution is probably that the tiny
lags have been hidden just inside the random error bars or their
magnitudes are smaller than the sensitivity limits of the detectors,
which leads us to conclude the lags in SGRBs are so insensitive to
the measurement that they can be neglected in statistics.

\subsection{The distribution of $\tau_{rel, 31}$}

It has been indicated that the widths and spectral lags of SGRBs in
comparison with those of long ones are usually smaller, however,
whether or not their ratios also follow the same law is an important
issue. Without losing generality, we choose the RSL $\tau_{rel, 31}$
as the $\tau_{rel}$ for this study. In answer to this question, we
make a plot of this distribution in figure 3,
\begin{figure}
\centering
   \includegraphics[width=4.2in,angle=0]{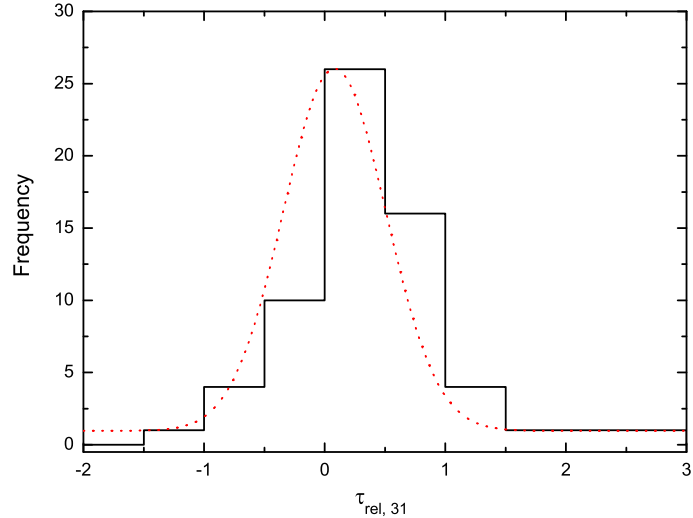}
  \caption{Distribution of RSL ($\tau_{rel, 31}$) for a sample of 65 single peaked SGRBs.
    The dotted line represents the best fit to the distribution with a gaussian function, where the expectation value
     and the standard deviation are $\mu=0.082$ and $\sigma=0.42$ respectively.}
  \label{Fig4}
\end{figure}
where we find the RSLs are normally distributed with
$\chi^{2}/dof=1.06$ and $R^2=0.99$. The expectation value and the
standard deviation are respectively $\mu=0.082$ and $\sigma=0.42$.

SGRBs behave similarly to long ones (see paper I) in the
distribution of RSLs, except that they differ in aspects of
deviation errors and expectation values. In evidence, the
discrepancy in the two classes originates directly from the fact
that unlike SGRBs most long bursts exhibit ``hard-to-soft''
evolution with positive spectral lags. Moreover, we notice that the
standard deviation of SGRBs is about 10 times larger than for long
bursts. If we take the $2\sigma$ sample RSL error as our confidence
limit, we can then estimate the effective upper limit to be
$2\sigma/\sqrt{n-1}\sim0.05$, which is reasonably close to the
value, 0.045, for the long burst sample. In principle, RSLs should
be normally distributed around zero if only the width of SGRB pulses
can be accurately measured. As shown in figure 3, the RSL
distribution indeed exhibits the trend although there is slight
departure between theory and observation due to some calculation
errors.

\section{Conclusion and discussion}

Our main conclusions in this work are as follows: Firstly, spectral
lags of SGRBs are normally distributed around zero, which causes
many of them to be negligible or unobservable; Secondly, the RSL
distribution is also found to be gaussian with the expectation value
of 0.082.

According to the internal shock model, GRBs are generally produced
by multiple relativistic shells (or wind) followed by internal
shocks due to their collision as a central engine pumps energy into
medium (see e.g. Fenimore et al. 1993; Rees \& M\'{e}sz\'{a}ros.
1994). It has been shown that the bulk Lorentz factor increases
linearly with radius until $\Gamma\leq 1000$ (Woods et al. 1995;
Ramirez-Ruiz \& Fenimore 2000; Eichler et al., 2000). Subsequently,
the motion of the shocked ejecta would be decelerated by the action
of the circumstellar medium (Huang et al. 1998, 2000; Piran 2005;
Granot \& Kumar 2006). However, the existence of discrete emission
regions with incompatible velocities causes the collisions between
them to be more intense and random. Because of the collisions of
these shells, the forward shock and the reverse shock could be
expected to occur at some distance from the central source, for
instance $R\sim10^{12-14} cm$ (Piran 1999), and propagate into the
relativistic ejecta. Under these conditions, SGRBs with negligible
spectral lags (due to large Lorentz factors) might take place.
Norris \& Bonnell (2006) had thought the high Lorentz factor,
$\Gamma\sim500-1000$, could interpret the phenomenon of negligible
lags for SGRBs. If the dependence of spectral lags, $\tau$, on
Lorentz factor, $\Gamma$, can be also expressed as
$\tau\propto\pm\Gamma^{-\omega}$ as in paper I, we can infer that
either high Lorentz factors or large $\omega$ (e.g. $\omega\gg2$),
or both of them, could lead to smaller lags towards zero. In a word,
the high Lorentz factor seems to be essential to this phenomenology.

Even though the statistical fluctuations and the spectral lags in
SGRBs can produce the same normal distribution, the
indistinguishable distributions couldn't exclude the smaller lags in
existence (Villasenor et al. 2005) if only lower energy bands (e.g.
channel 1: 20-50 Kev) are taken into account (Norris \& Bonnell
2006), since the contribution of curvature effects to spectral lags
in this case would cause longer decays in pulses, and thus larger
time lags. However, the influence of angular-spreading effects on
the decay phase of pulses would be reduced with the decrease of
radius of emission region. When the radii reduce to a certain limit,
the curvature effects can then be eliminated. Under the
circumstances, the time lags in GRBs should be unobservable. We
hence deduce that SGRBs might locate at the smaller distance from
the central engine.

Our findings in this paper are based on the previous observations of
single pulse SGRBs. The conclusions need to be further verified by
much more precise observations with the ongoing satellites HETE-2
and SWIFT, which would shed new light on the nature of spectral lags
in SGRBs, even their physical mechanism and origin.

\section*{Acknowledgments}
We acknowledge the anonymous referee's helpful and constructive
comments. We are also thankful to Dr R. S. Pokorny for his
invaluable help.
\appendix

\clearpage

\label{lastpage}

\end{document}